# First tests of "bulk" MICROMEGAS with resistive cathode mesh

R. Olivera,[1] V. Peskov,[1,2] Pietropaolo,[3] P.Picchi[4]
[1]CERN, Geneva, Switzerland
[2]UNAM, Mexico
[3]INFN Padova, Padova, Italy
[4]INFN Frascati, Frascati, Italy

**Abstract**
We present the first results from tests of a MICROMEGAS detector manufactured using the so- called "bulk" technology and having a resistive cathode mesh instead of the conventional metallic one. This detector operates as usual MICROMEGAS, but in the case of sparks, which may appear at high gas gains, the resistive mesh reduces their current and makes the sparks harmless. This approach could be complementary to the ongoing efforts of various groups to develop spark-protected MICROMEGAS with resistive anode planes.

## 1. Introduction

During the last year there have been intense efforts from various groups to develop spark –protected micropattern gaseous detectors with resistive electrodes (see for example [1] and references there in). In our recent report [2] we presented the first results from tests of "home- made" MICROMEGAS with a resistive cathode mesh. As was shown earlier, resistive electrodes may reduce the discharge current by a factor of 1000 and thus protect the detector and the front end electronics from being damaged. We named this detector a Resistive Mesh MICROMEGAS or RM-MICROMEGAS.
Home-made designs of RM-MICROMEGAS had 50 or 100 μm width spacers made of Nylon fishing lines or from Kapton meshes and were very convenient for general studies and design optimization since they allow easy changes to be made. It is obvious, however, that this primitive detector assembling technique has a lot of limitations. Moreover it was observed that the maximum achievable gain depends on the spacer shape and materials from which they were made.
  In this report we present preliminary results from the tests of the first prototype RM-MICROMEGAS with resistive cathode mesh manufactured by a so-called "bulk" technology [3]. With the bulk RM-MICROMEGAS we were able to achieve almost the same gas gains as with conventional MICROMEGAS having a metallic cathode mesh. The main advantages of this technology is that it has the potential to build large-area, high quality, spark –protected detectors.

## 2. Materials and methods

The resistive meshes used for the construction of RM-MICROMEGAS were manufactured by a laser drilling technique from the resistive Kapton 100XC10E5 (resistivity 2.8-3 MΩ/□) used earlier by us in the RETGEM designs [4]. The resistive meshes had a thickness of 20μm, a hole diameter d=50 μm and hole spacing a=100 μm. The bulk RM-MICROMEGAS was manufactured in a few steps. First the anode plate (fiber glass with a 20μm thick Cu layer) and a photosensitive film (DUPONT photoimageable coverlay PC1025) having the right thickness and the resistive mesh were laminated together at an elevated temperature forming a monolithic object. The photosensitive film was then photolitographically etched so that supporting pillars were formed. These pillars had a cylindrical shape with a 300μm diameter, their height was 128μm and the pitchwas 2mm (see Fig.1).
It is known that the "bulk" technology allows large-area MICROMEGAS to be built, however for the sake of simplicity the first prototype had an active area of only 5x5cm$^2$ (see Fig. 2). The anode plate was manufactured from a Cu coated fiber glass printed circuit plate.
The RM-MICROMEGAS was tested in a gas chamber using an experimental setup shown in Fig.3. In comparative studies we also used several conventional MICROMEGAS including one manufactured by the bulk technology.

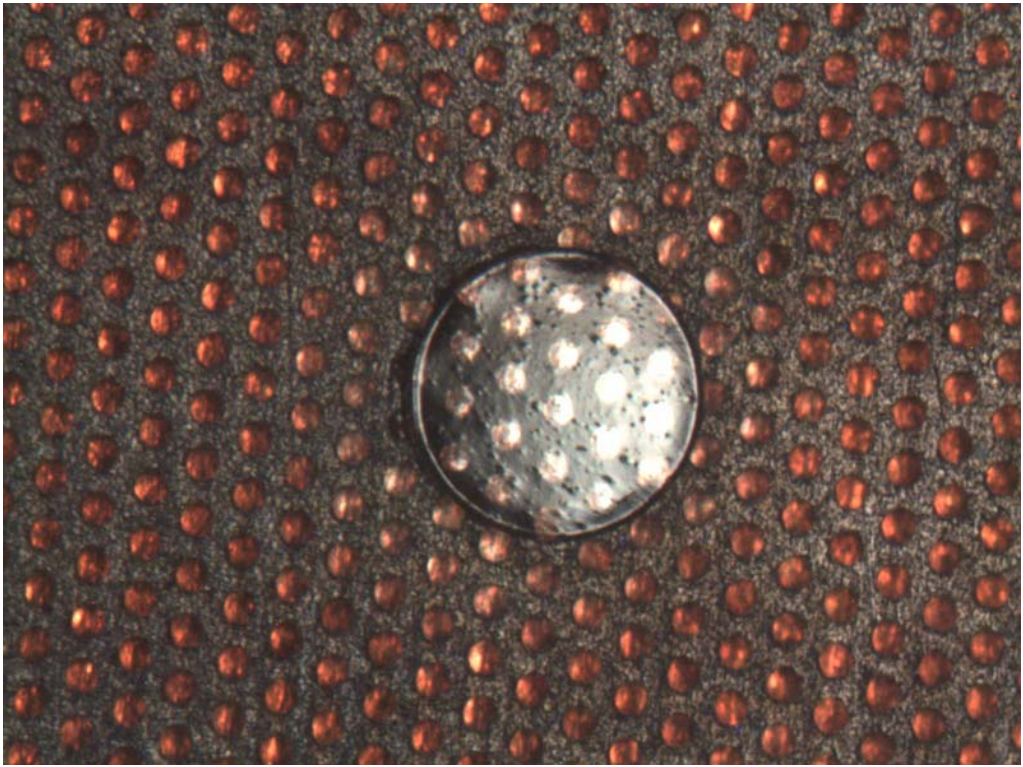

Fig. 1. A magnified photograph of the bulk MICROMEGAS cathode resistive mesh with incorporated supporting pillars under it.

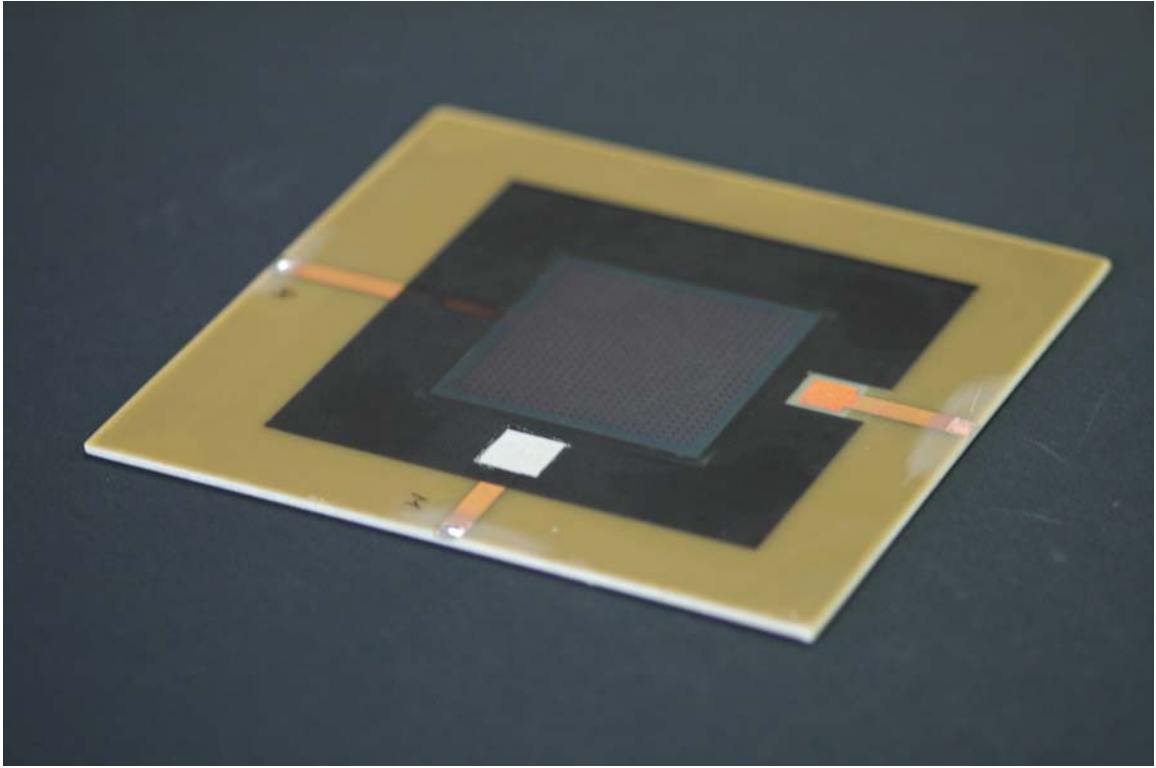

Fig. 2. A photograph of the bulk MICROMEGAS with resistive mesh. Regularly distributed dots over the sensitive area are the mesh supporting pillars.

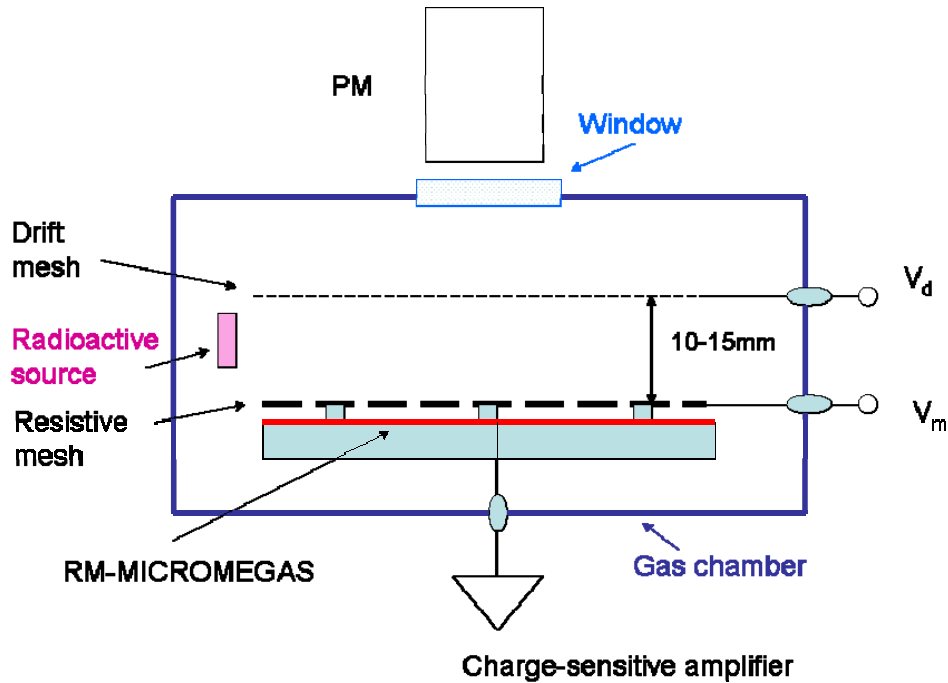

Fig. 3. A schematic drawing of the experimental setup used for the RM-MICROMEGAS studies

The chamber was flushed with one of the following gas mixtures: Ar+5%isobutane, Ne+7%$CH_4$ or Ar+7%$CH_4$.
The primary ionization in the drift region of the detector was produced either by alpha particles from $^{241}$Am or by 6keV photons from $^{55}$Fe. The avalanche signals were detected by a charge sensitive amplifier Ortec 142pc. Besides the maximum achievable gain measurements it was also very important to evaluate the spark protective properties of the resistive mesh. For this we used the same electronic circuit as in [5] in which was compared the charge deposited by sparks in a conventional MICROMEGAS and in a MICROMEGAS with the anode coated by resistive Kapton. This allows us to correctly compare our results obtained with the resistive cathode mesh with their results obtained with the resistive anode.[*]
In complementary measurements the spark energy was also independently evaluated by its light emission; for this a photomultiplier EMI-9426 was used (see Fig.3)

---

[*] Note that in the past in similar comparative studies performed with TGEMs and RETGEMs we used a current amplifier with a changeable feedback resistor which allowed us to measure the spark current values in a large interval. In the case of measuring high currents the low resistor should be used and the energy released by the spark was so high that it could damage the metallic mesh of the conventional MICROMEGAS. Thus the circuit shown in fig.4 was more adequate for these particular measurements.

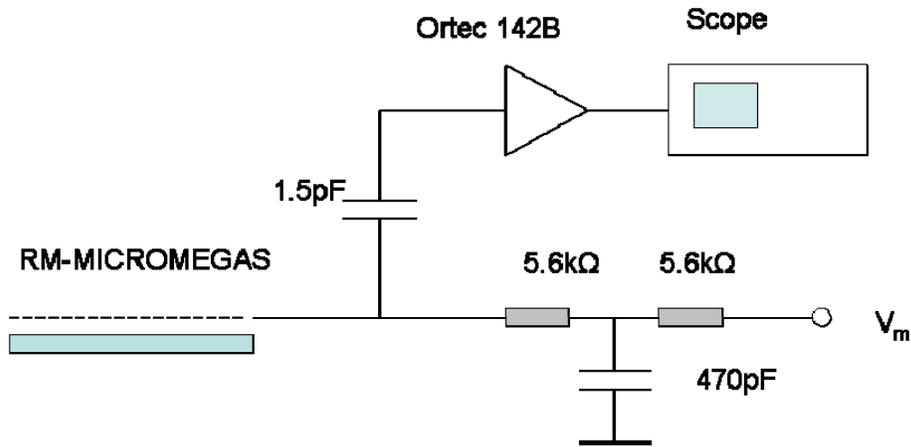

Fig. 4. A schematic of the circuit used for measurements of the charge released by sparks (see [5] for more details).

### 3. Preliminary results

Most of the tests were done in Ar+5%isobutane. This mixture was chosen in order to compare our results with the earlier published data for bulk MICROMEGAS with a metallic mesh(see for example [3]).
Fig.5 shows gain vs. voltage curves measured in Ar+isobutane with the RM-MICROMEGAS and with the conventional bulk MICROMEGAS using $^{241}$Am and $^{55}$Fe radioactive sources. In the case of the Am source we were able to observe signals in ionization chamber mode (at gas gain of one). With the voltage increase the multiplication starts and the gas gain as a function of the voltage V was defined as:
$$A(V)= S(V)_g/S(V)_i,$$
where $S_g(V)$ and $S_i(V)$ are the signal amplitudes measured in the case of gas multiplication and in the ionization chamber mode respectively.
At gas gains of more than hundred $^{55}$Fe was used. The gas gain in this case was evaluated from the known sensitivity of the preamplifier (calibrated by a standard method of the charge injection into the preamplifier/detector input). The highest points on the gas gain curves measured with $^{55}$Fe corresponds to the situation when the first signs of detector instability appear. Usually at ~20V higher voltage breakdowns may appear at a rate of about a few per 10 min. Their rate then sharply increases with voltage.
As can be seen from the curves presented in Figure 5, the maximum gas gain (~$10^4$) achieved with RM-MICROMEGAS is only two times less then that measured with the conventional MICROMEGAS; the last one has the maximum achievable gain comparable to one published in [3].
Note that the maximum achievable gain of RM-MICROMEGAS may depend on the gas mixture. For example, Fig. 6 depicts gain vs. voltage curves measured in Ne+7%CH$_4$ and inAr+7%CH$_4$; as can be seen the maximum achievable gain in Ar+7%CH$_4$ is about a factor of 10 less than in Ar+5%isobutae or in Ne+7%CH$_4$.

The energy resolution measured when tall the detector area was illuminated by $^{55}$Fe was about 30%FWHM (see Fig.7).

The rate characteristics of the RM-MICROMEGAS (the signal amplitudes vs. the counting rate) are presented in Figure 8. As can be seen no significant reduction in pulse amplitude was observed up to counting rates of $10^4$Hz/cm$^2$.

The results of the measurements of the charge deposited by sparks in RM-MICROMEGAS and in the conventional MICROMEGAS were very similar to those obtained in [5]: the charge released by sparks in RM-MICROMEGAS was typically about ~$10^{10}$ electrons which is at least 100 times less than in MICROMEGAS with a metallic cathode mesh (usually $\geq 10^{12}$ electrons). Qualitatively this was confirmed by measurements performed with the photomultiplier: typically the spark light emission in RM-MICROMEGAS was at least 10 times less than in the conventional one. Thus RM-MICROMGAS can operate at gas gains as high as the conventional MICROMEGAS. However in the case of sparks the resistive mesh considerably reduces their energy, making the detector spark-protected.

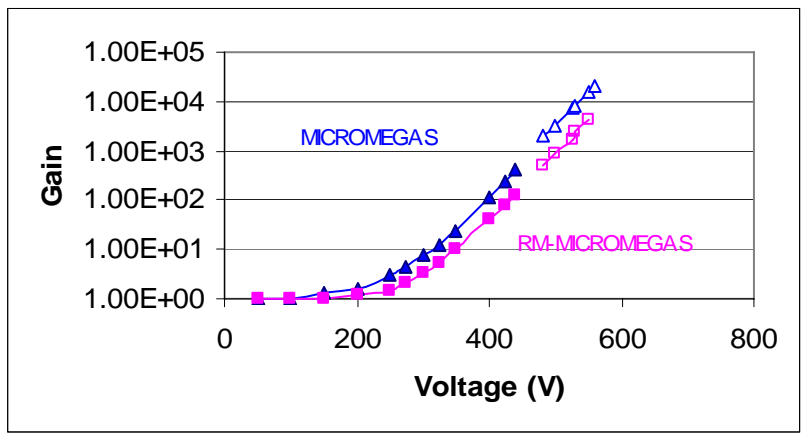

Fig.5. Gain vs. voltage curves measured in Ar+5% isobutene with bulk RM-MICROMEGAS (squares) and with the conventional bulk MICROMEGAS (triangles). Filled symbols correspond to measurements performed with alpha particles ($^{241}$Am), open symbols- 6keV photons from $^{55}$Fe.

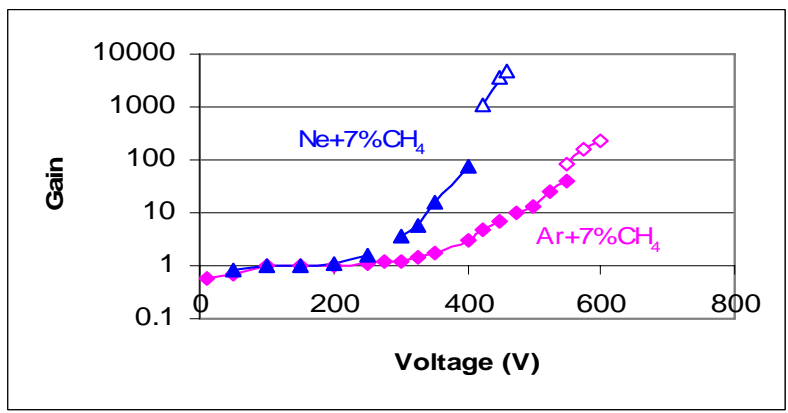

Fig. 6. Gain curves for RM-MICROMEGAS operating in Ne +7%CH$_4$ (triangles) and in Ar+7%CH$_4$. Filled symbols - $^{241}$Am, open symbols - $^{55}$Fe.

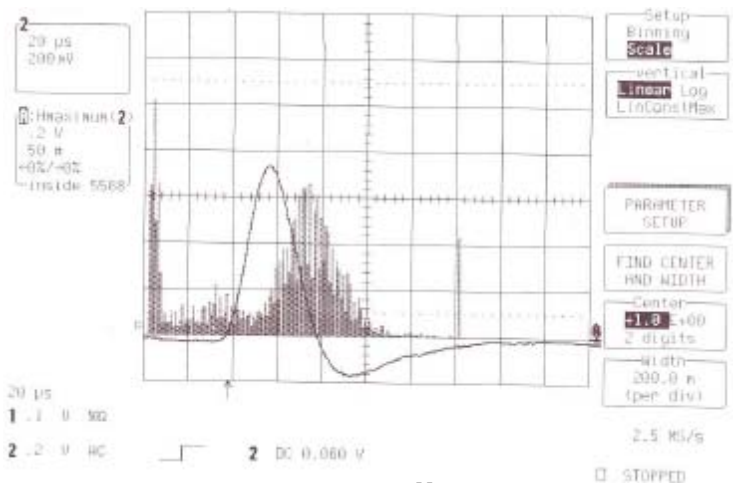

Fig.7. Pulse-height spectrum of $^{55}$Fe measured with RM-MICROMEGAS operating in Ar+5%isobutane at a gas gain of $810^3$ in Ar+5%isobutane

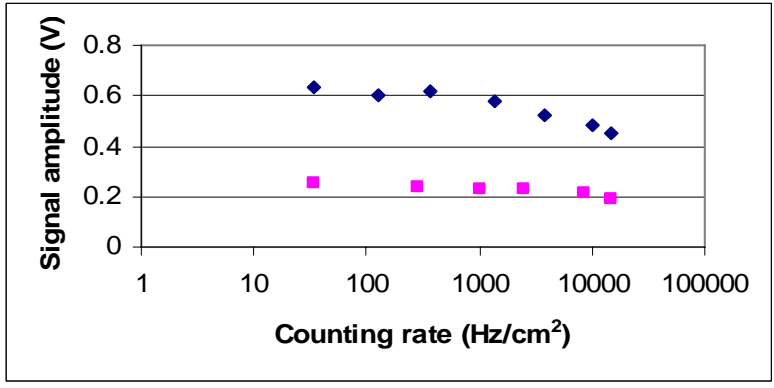

Fig.8. Signal amplitude vs. the counting rate measured for the RM-MICROMEGAS at a gas gain of $\approx 810^3$ (rhombus) and $\approx 2.510^3$ (squares).

**Conclusions**

Successful implementation of the bulk technology for RM-MICROMEGAS manufacturing demonstrates the possibility of building high quality, large- area, spark-protected detectors of photons and charged particles by an industrial process. This approach could be complementary to ongoing efforts performed by various groups within the framework of the RD51 collaboration to develop MICROMEGAS and other types of gaseous detectors with the resistive electrodes.